\documentclass[12pt,a4paper]{article}
\usepackage{epsfig}
\usepackage[nooneline,flushleft]{caption}
\usepackage{subfigure}
\usepackage{amsmath}
\newcommand{\spp}{\vphantom{\bigg(}}
\newcommand{\ord}{{\cal O}}

\def\bqll{B_s \rightarrow l^+ \, l^-}

\def\gev{{\rm GeV}}

\begin{document}
\title{Rare decays $B_s\to l^+l^-$ and $B\to Kl^+l^-$ in \\the
topcolor-assisted technicolor model
\\
\hspace*{-0.8cm}  }

\author{Wei Liu, Chong-Xing Yue, Hui-Di Yang\\
{\small  Department of Physics, Liaoning Normal University, Dalian
116029, China}\thanks{cxyue@lnnu.edu.cn}\\}
\date{}

\maketitle
\begin{abstract}
We examine the rare decays $B_s\to l^+l^-$ and $B\to Kl^+l^-$ in the
framework of  the topcolor-assisted technicolor ($TC2$) model. The
contributions of the new particles predicted by this model to these
rare decay processes are evaluated. We find that the values of their
branching ratios are larger than the standard model predictions by
one order of magnitude in wide range of the parameter space. The
longitudinal polarization asymmetry of leptons in $B_s \to l^+l^-$
can approach $\ord(10^{-2})$. The forward-backward asymmetry of
leptons in $B \to Kl^+l^-$ is not large enough to be measured in
future experiments. We also give some discussions about the
branching ratios and the asymmetry observables related to these rare
decay processes in the littlest Higgs model with T-parity.
\end{abstract}

\newpage

\noindent{\bf \large I. Introduction}

The study of pure leptonic and semileptonic decays of B meson is one
of the most important tasks of B physics both theoretically and
experimentally. These rare B decays are sensitive to new physics
($NP$) and their signals are useful for testing the standard model
($SM$) \cite{bdk}. So far, a lot of works have been concentrated on
these decays. In the $SM$, there are no flavor changing neutral
current ($FCNC$) processes at the tree level and the leading
contributions to these decays come from the one-loop level. So these
rare decays are rather sensitive to the contributions from the $NP$
models beyond the $SM$. Studying of the observables of the
asymmetries, such as the $CP$ asymmetry \cite{cp}, longitudinal
polarization ($LP$) asymmetry $A_{LP}$ \cite{lp}, and
forward-backward ($FB$) asymmetry $A_{FB}$ \cite{fb} etc, interests
experiments in testing $NP$. Certainly, their detection requires
excellent triggering and identification of leptons with low
misidentification rates for hadrons. The precision measurement needs
further studying.

The quark level transition $b \to sl^+l^-$ is responsible for both
the purely leptonic decays $B_s \to l^+l^-$ and the semileptonic
decays $B \to Kl^+l^-$($l=e,\mu,\tau$). The decay $B_s \to
\mu^+\mu^-$ will be one of the most important rare B decays to be
studied at the upcoming large hadron collider ($LHC$), and so far
the upper bound on its branching ratio is \cite{exp1}
\begin{equation}
Br(B_s \to \mu^+\mu^-)<5.8\times10^{-8}(95\% ~{\rm
C.L.}).\label{exp111}
\end{equation}
The branching ratios of $B\to K l^{+} l^{-}$ observed by BaBar
collaboration and Belle collaboration are \cite{babar-04,exp2}
\begin{eqnarray}
Br(B\rightarrow K l^{+} l^{-})  = (5.7^{+2.2}_{-1.8})\times10^{-7},
\label{exp222}
\end{eqnarray}
which is close to the $SM$ prediction \cite{bdk,lunghi}. However,
due to the errors in the determination of the hadronic form factors
and the Cabibbo-Kobayashi-Maskawa ($CKM$) matrix element $|V_{ts}|$,
there is about $20 \%$ uncertainty in $SM$ prediction. The
experimental measurement values of rare decay processes $B_s\to e
^+e^-,\tau^+\tau^-$ will be discussed later.

We also consider other observables of the purely leptonic and
semileptonic decays for the $B$ meson, which are sensitive to
scalar/pesudoscalar new physics ($SPNP$) contributions to $b \to s$
transitions. They are forward-backward asymmetry $A_{FB}$ of leptons
in $B \to Kl^+l^-$ and longitudinal polarization asymmetry $A_{LP}$
of leptons in $B_s \to l^+l^-$. The observable $A_{LP}$ was
introduced in Reference \cite{lp}, though the corresponding analysis
in the context of $K \to \mu^+\mu^-$ had been carried out earlier
\cite{kmu}. The average $A_{FB}$ in the rare decay processes $B \to
Kl^+l^-$ has been measured by BaBar collaboration as \cite{babar-04}
\begin{equation}
\left\langle A_{FB}\right\rangle  =  0.15_{-0.23}^{+0.21} \pm 0.08.
\end{equation}
This measured value is close to zero and has a high experimental
error. As the values of $A_{LP}$ and $A_{FB}$ predicted in the $SM$
are nearly zero, any nonzero value of one of these asymmetries is a
signal for $NP$. This is the main reason we focus on these
observables.

In literature, there are numerous studies of the quark level decays
$b \to sl^+l^-$ both in the $SM$ and in some $NP$ models. Recently,
Reference \cite{ued,ued1} have studied the sensitivity of these rare
decay processes to the radius $R$ in the universal extra dimension
($UED$) model. In the supersymmetry ($SUSY$) models, extensive works
have been taken to the branching ratios of these rare decays, and
some of these discussions are related to the asymmetry aspect
\cite{yang,susy}. These decays have also been discussed in the
littlest Higgs model with T-parity (called the $LHT$ model)
\cite{6}, they have verified that the $LHT$ model can enhance the
branching ratios of these decays \cite{lht}. However, they have not
discussed the asymmetry observables, we will give some discussions
on these observables in the framework of the $LHT$ model.

In the framework of the topcolor-assisted technicolor ($TC2$) model
\cite{tc209}, Reference \cite{tc201} has calculated the branching
ratios of quark level $b \to sl^+l^-$ decays. They consider the
contributions of the non-universal gauge boson $Z'$ predicted by
this model. Their numerical results show that the enhancement is
quite large when the mass of $Z'$ is small. Reference \cite{tc202}
has calculated the contributions coming from the pseudoscalar
top-pions predicted by this model to the branching ratios of the
decays $B_s \to l^+l^-$. Reference \cite{xiong} has evaluated the
contributions from both the neutral and charged scalars predicted by
this model, the branching ratios can be enhanced over the $SM$
predictions by two orders of magnitude in some part of parameter
space. So far, we have not seen the study of the asymmetry
observables for these two decays in the framework of the $TC2$
model, and furthermore the former discussions on the branching
ratios have not considered the contributions induced by all the
particles predicted by this model.

In this paper, we consider the contributions coming from all of the
new particles predicted by the $TC2$ model to the branching ratios
and asymmetries related to the rare decay processes $b \to sl^+l^-$.
Compared with the predictions in the $SM$, our results show that the
contributions to the branching ratios and the asymmetries come from
two aspects. First, the Wilson coefficients of these processes
receive additional contributions from the non-universal gauge boson
$Z'$ and charged top-pions. Second, the neutral top-pion and
top-Higgs can give contributions through newly introduced
scalar/pesudoscalar operators. For comparison, we also give our
results in the $LHT$ model, considering different parametrization
scenarios.

This paper is arranged as follows. In the following section, we will
summarize some elementary features of the $TC2$ model. In Sec. III
we present our calculation on the decay processes $B_s \to l^+l^-$.
The decay processes $B \to Kl^+l^-$ will be studied in the Sec. IV.
In Sec. V we give simple discussions on the above questions in the
$LHT$ model. Conclusions are given in Sec. VI.

\noindent{\bf \large II. The $\rm{TC2}$ model}

The $TC2$ model \cite{tc209} is one kind of the phenomenological
viable models, which has all essential features of the topcolor
scenario. The $TC2$ model generates the large quark mass through the
formation of a dynamical $t\bar t$ condensation and provides
possible dynamical mechanism for  electroweak symmetry breaking
($EWSB$). The physical top-pions($\pi_t^{0,\pm}$), the non-universal
gauge boson ($Z'$) and the top-Higgs ($h_t^0$) are predicted. The
presence of the physical top-pions $\pi_{t}^{0,\pm}$ in the low
energy spectrum is an inevitable feature of the topcolor scenario,
regardless of the dynamics responsible for $EWSB$ and other quark
mass. The flavor-diagonal ($FD$) couplings of top-pions to fermions
can be written as \cite{tc209,tc203}:
\begin{eqnarray}
&&\frac{m_t^*}{\sqrt{2}F_{\pi}} \frac{\sqrt{\nu_{w}^{2}-
F_{\pi}^{2}}}{\nu_{w}}\left[i\bar{t}\gamma^{5}t\pi_{t}^{0}+\sqrt{2}\bar{t}_{R}b_{L}
\pi_{t}^{+}+\sqrt{2}\bar{b}_{L}t_{R}\pi_{t}^{-}\right] \nonumber \\
&&+\frac{m_{b}^{*}}{\sqrt{2}F_{\pi}}\left[i\bar{b}\gamma^{5}b\pi_{t}^{0}+\sqrt{2}
\bar{t}_{L}b_{R}\pi_{t}^{+}+\sqrt{2}\bar{b}_{R}t_{L}\pi_{t}^{-}\right]
+\frac{m_{l}}{\nu}\bar{l}\gamma^{5}l\pi^{0}_{t},
\end{eqnarray}
where $m_t^*=m_{t}(1-\varepsilon)$, $\nu_{w}=\nu/\sqrt{2}=174\rm
~GeV$, $F_{\pi}\approx50\rm ~GeV$ is the top-pion decay constant.
The ETC interactions give rise to the masses of the ordinary
fermions including a very small portion of the top quark mass,
namely $\varepsilon m_{t}$ with a model dependent parameter
$\varepsilon\ll 1$, and $m_b^*=m_b-0.1\varepsilon m_t$ \cite{jpg}.
The factor $\frac{\sqrt{\nu_{w}^{2}-F_{\pi}^{2}}}{\nu_{w}}$ reflects
mixing effect between top-pions and the Goldstone bosons.

For the $TC2$ model, the underlying interactions, topcolor
interactions, are non-universal and therefore do not posses
Glashow-Iliopoulos-Maiani ($GIM$) mechanism \cite{gim}. One of the
most interesting features of $\pi_{t}^{0,\pm}$ is that they have
large Yukawa couplings to the third-generation quarks and can induce
the tree-level flavor changing ($FC$) couplings \cite{tc204,y4}.
When one writes the non-universal interactions in the quark mass
eigen-basis, it can induce the tree-level $FC$ couplings. The $FC$
couplings of top-pions to quarks can be written as
\cite{tc201,tc204}:
\begin{eqnarray}
&&\frac{m_{t}^*}{\sqrt{2}F_{\pi}}\frac{\sqrt{\nu_{w}^{2}-
F_{\pi}^{2}}}{\nu_{w}}\left[iK_{UR}^{tc}K_{UL}^{tt^{*}}\bar{t}_{L}c_{R}\pi_{t}^{0}
+\sqrt{2}K_{UR}^{tc^{*}}K_{DL}^{bb}\bar{c}_{R}b_{L}\pi_{t}^{+}
+\sqrt{2}K_{UR}^{tc}K_{DL}^{bb^{*}}\bar{b}_{L}c_{R}\pi_{t}^{-}\right.\nonumber \\
&&\hspace{15mm}\left.+\sqrt{2}K_{UR}^{tc^{*}}K_{DL}^{ss}\bar{t}_{R}s_{L}\pi_{t}^{+}
+\sqrt{2}K_{UR}^{tc}K_{DL}^{ss^{*}}\bar{s}_{L}t_{R}\pi_{t}^{-}\right]
,
\end{eqnarray}
where $K_{UL(R)}$ and $K_{DL(R)}$ are rotation matrices that
diagonalize the up-quark  and down-quark mass matrices $M_{U}$ and
$M_{D}$, i.e., $K_{UL}^{+} M_{U}K_{UR}=M_{U}^{dia}$ and
$K_{DL}^{+}M_{D}K_{DR}=M_{D}^{dia}$, for which the $CKM$ matrix is
defined as $V=K_{UL}^{+}K_{DL}$. To yield a realistic form of the
$CKM$ matrix $V$, it has been shown that the values of the coupling
parameters can be taken as \cite{tc204}:
\begin{equation}
K_{UL}^{tt}\approx K_{DL}^{bb} \approx K_{DL}^{ss}\approx1,
\hspace{10mm} K_{UR}^{tc}\leq\sqrt{2\varepsilon-\varepsilon^{2}} .
\end{equation}
In the following calculation, we will take
$K_{UR}^{tc}=\sqrt{2\varepsilon- \varepsilon^{2}}$ and take
$\varepsilon$ as in the range of $0.03-0.1$ \cite{tc209}. The $TC2$
model predicts the existence of the top-Higgs $h_t^0$, which is a
$t\bar t$ bound and analogous to the $\sigma $ particle in low
energy $QCD$. It has similar Feynman rules as the $SM$ Higgs boson,
so we don't list them.

Another significant feature of the $TC2$ model is the existence of
non-universal gauge boson $Z'$, which may provide significant
contributions to some $FCNC$ processes because of its $FC$ couplings
to fermions. The $FC$ $b-s$ coupling to $Z'$ can be written as
\cite{tc210}:
\begin{eqnarray}
  {\cal L}^{FC}_{Z'}=-\frac{g_1}{2}\cot{\theta'} Z'^{\mu}\left\{\frac{1}{3}D_{L}^{bb} D_{L}^{bs*}
  \bar{s}_L\gamma_{\mu} b_L -
\frac{2}{3}D_{R}^{bb}D_{R}^{bs*}\bar{s}_R \gamma_{\mu} b_R  +{\rm
h.c.} \right\} \label{bsz},
\end{eqnarray}
$D_L,D_R$ are matrices which rotate the down-type left and right
hand quarks from the quark field to mass eigen-basis. The $FD$
couplings of $Z'$ to fermions, which are relative to our
calculation, can be written as \cite{tc209,tc201,tc203,tc211}:
\begin{eqnarray}
{\cal L}^{FD}_{Z'}&=&-\sqrt{4\pi K_{1}}\left\{
Z'_{\mu}\left[\frac{1}{2}\bar{\tau}_{L}\gamma^{\mu}\tau_{L}
-\bar{\tau}_{R}\gamma^{\mu}\tau_{R}+\frac{1}{6}\bar{t}_{L}\gamma^{\mu}t_{L}
+\frac{1}{6}\bar{b}_{L}\gamma^{\mu}b_{L}+\frac{2}{3}\bar{t}_{R}\gamma^{\mu}t_{R}
\right.\right.\nonumber\\
&-&\left.\left.\frac{1}{3}\bar{b}_{R}\gamma^{\mu}b_{R}\right]-\tan^{2}
\theta'Z'_{\mu}\left[\frac{1}{6}\bar{s}_{L}\gamma^{\mu}s_{L}
-\frac{1}{3}\bar{s}_{R}\gamma^{\mu}s_{R}-\frac{1}{2}
\bar{\mu}_{L}\gamma^{\mu}\mu_{L}-\bar{\mu}_{R}\gamma^{\mu}\mu_{R}
\right.\right.\nonumber\\
&-&\left.\left.\frac{1}{2}\bar{e}_{L}\gamma^{\mu}e_{L}-\bar{e}_{R}\gamma^{\mu}e_{R}\right]\right\},
\label{zll}\end{eqnarray} where $K_{1}$ is the coupling constant and
$\theta'$ is the mixing angle with $\tan
\theta'=\frac{g_{1}}{\sqrt{4\pi K_{1}}}$. $g_{1}$ is the ordinary
hypercharge gauge coupling constant.

In the following sections, we will use the above formulae to
calculate the contributions of the $TC2$ model to the rare decay
processes $B_s\to l^+l^-$ and $B\to Kl^+l^-$.

\noindent{\bf \large III. The contributions of the $\rm{TC2}$ model
to the rare decay processes \hspace*{0.7cm}$B_s \to l^+l^-$}

The $TC2$ model can give contributions to rare B decays two
different ways, either through the new contributions to the Wilson
coefficients or through the new scalar or pseudoscalar operators.
The most general model independent form of the effective Hamilton
for the decays $B_s \to l^+ l^-$ including the contributions of $NP$
has the form:
\begin{equation}
H(B_s \to l^+ l^-) = H_{0} + H_{1}\, \label{lag-tot}
\end{equation}
with
\begin{eqnarray}
H_{0}  &=&  \frac{ \alpha G_F}{2 \sqrt{2} \pi}
  \left( V_{ts}^* V_{tb} \right) \,
  \biggl\{
  R_A
  (\bar{s} \, \gamma_\mu \gamma_5 \, b)
  (\bar{l} \, \gamma^\mu \gamma_5 \, l)
  \biggr\} \;,
  \label{hsm}
\end{eqnarray}
\begin{equation}
H_{1}  =  \frac{\alpha G_F}{\sqrt{2} \pi}( V_{tb} V^*_{ts}) \biggl\{
R_S \, (\bar{s}\,P_R\, b) \, (\bar{l} \,l) + R_P \, (\bar{s}\,P_R\,
b) \, (\bar{l} \gamma_5 l) \biggr\} \; . \label{hsp}
\end{equation}
Where $H_0$ represents the $SM$ operators and $H_1$ represents the
$SPNP$ operators. Here $P_{L,R} = (1 \mp \gamma_5)/2$, $R_S$, $R_P$
and $R_A$ denote the strengths of the scalar, pseudoscalar, and
axial vector operators, respectively \cite{0805}. In our analysis we
assume that there are no additional $CP$ phases apart from the
single $CKM$ phase, thus $R_S$ and $R_P$ are real. In the $SM$, the
scalar and pseudoscalar couplings $R_S$ and $R_P$ receive
contributions from the penguin diagrams with physical and unphysical
neutral scalar exchange and are highly suppressed to
$\ord(10^{-5})$. The coupling constant of the axial vector operator
$R_A$ can be expressed as $R_A = {Y^{SM}(x)}/{\sin^2 \theta_w}$,
where $Y^{SM}(x)$ is the $SM$ Inami-Lim function \cite{smf}, which
has been listed in Appendix A. These coupling constants will receive
contributions coming from the non-universal gauge boson $Z'$ and the
scalars $\pi_t^{0,\pm},h^0_t$.

\noindent{ \bf A. The contributions of the nonuniversal gauge boson
$Z'$}

\hspace{1cm}
\begin{figure}[htb]
\begin{center}
\epsfig{file=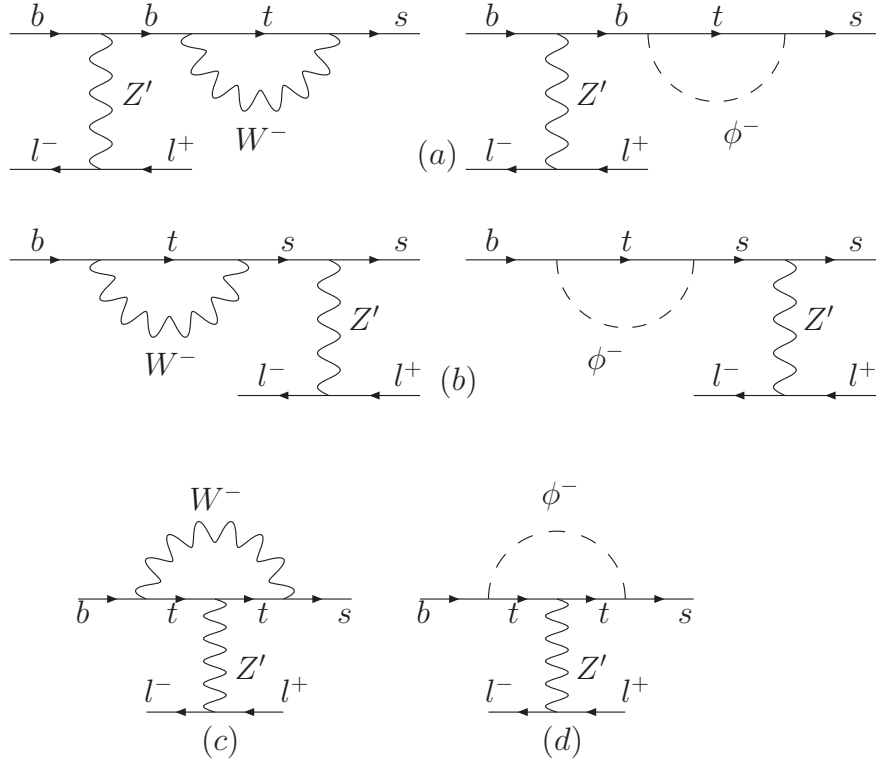,scale=0.85} \caption{Penguin diagrams of $Z'$
contributing to $B_s\to l^+l^- $ in the $TC2$ model.} \label{a1}
\end{center}
\end{figure}
\hspace{1cm}

In the TC2 model, the non-universal gauge boson $Z'$ can give
corrections to the $SM$ function $Y(x)$, which directly determine
the coupling constant $R_A$. The relevant Feynman diagrams have been
shown in Fig.\ref{a1}. In these diagrams, the Goldstone boson $\phi$
is introduced by the 't Hooft-Feynman gauge, which can cancel the
divergence in self-energy diagrams. Because the couplings of $Z'WW$,
$Z'\phi\phi$ and $Z'W\phi$ do not exist in the $TC2$ model, the
diagrams that including the above couplings are not present. The
small interference effects between $Z'$ and $Z$ are not considered
here. In this situation, the function $Y^{TC}(x_t)$ for $l=e, \mu$
is obtained as follows:
\begin{eqnarray}
Y^{TC}(x_t)&=&\frac{-tan^2\theta'
M_Z^2}{M_{Z'}^2}\left(C_{ab}(x_t)+C_c(x_t)+C_d(x_t)\right)\label{ytcz},
\end{eqnarray}
here $x_t=m_t^{*2}/M_W^2$. The factor $-tan^2\theta'$ does not exist
for the decay process $B_s\to \tau^+\tau^- $ which can be seen from
Eq.~(\ref{zll}). The formations of $C_{ab}(x_t)$, $C_c(x_t)$ and
$C_d(x_t)$ can be easily obtained in the framework of the $TC2$
model using the method in Reference \cite{smf}. The detailed
expression forms of these functions are listed in the Appendix B.

\begin{figure}[htb]
\begin{center}
\epsfig{file=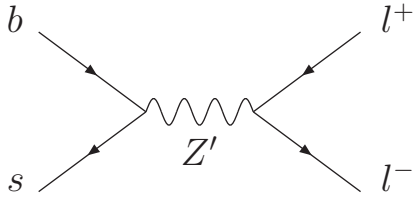,scale=1.0} \caption{Tree level diagram of $Z'$
contributing to $B_s\to l^+l^- $ within the $TC2$ model.} \label{a3}
\end{center}
\end{figure}

The non-universal gauge boson $Z'$ has $FC$ coupling with fermions
as shown in Eq.~(\ref{bsz}), the tree level Feynman diagram
contributing to the decay processes $B_s\to l^+l^- $ has been shown
in Fig.\ref{a3}. The contributions can be obtained by directly
calculating Fig.\ref{a3} using the standard method in Reference
\cite{tc210}, and the $B_s$ width can be written as:
\begin{equation}
\Gamma(B_s\to l^+l^-)=\frac{1}{4608\,\pi} f_{B_s}^{2}
m_{B_s}m_{l}^{2}\,\sqrt{1- \frac{4m_{l}^{2}}{m_{B_s}^{2}}}
\,\delta_{bs}^{2}\cot^2\theta ' X^2(\theta ')
\left(\frac{g_1}{M_{Z'}}\right)^4 ,\label{gamma_ll}
\end{equation}
where
\begin{equation}
\delta_{bs}=D_{L}^{bb}D_{L}^{bs*} +2D_{R}^{bb}D_{R}^{bs*}.
\label{delta}
\end{equation}
$X(\theta ')=\cot\theta '$ for $l=\tau$, and $X(\theta ')=\tan\theta
'$ for $l=e$ and $\mu$. $f_{B_s}$ is the decay constant of $B_s$
meson.

\noindent{ \bf B. The contributions of the scalars
($\pi_t^{0,\pm},h^0_t$)}

\hspace{0.5cm}
\begin{figure}[htb]
\begin{center}
\epsfig{file=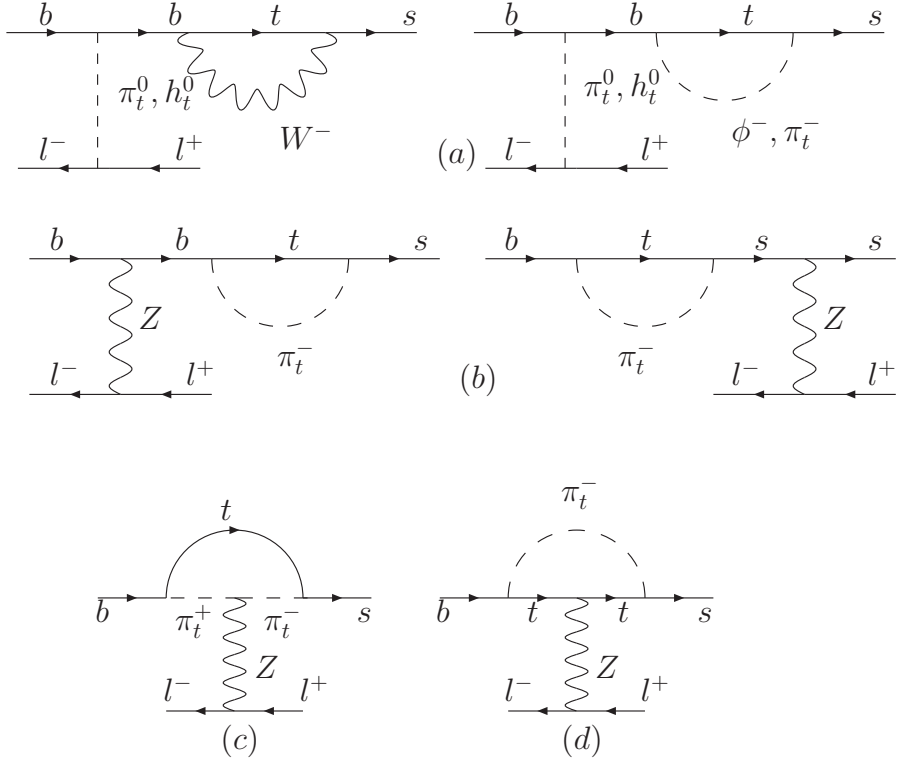,scale=0.85} \caption{Scalar particles
contributing to $B_s\to l^+l^- $ in the $TC2$ model.} \label{a2}
\end{center}
\end{figure}

The scalars predicted by the $TC2$ model give contributions to the
decay processes $B_s \to l^+l^-$ through corrections to the coupling
constants in Eq.~(\ref{hsm}) and Eq.~(\ref{hsp}). The relevant
Feynman diagrams are displayed in Fig.\ref{a2}, in which $(a)$ shows
the contributions of neutral top-Higgs $h^0_t$ and top-pion
$\pi_t^0$ to the couplings $R_S$ and $R_P$, respectively; $(b)$, $
(c)$ and $ (d)$ show the contributions of the charged top-pions
$\pi_t^{\pm}$ to the coupling $R_A$. The expression of the
coefficient $R_S$ can be written as:
\begin{equation}
R_{S}  = \frac{\sqrt{\nu_w^2-F_{\pi}^2}}{\nu_w}\left(
\frac{m_b^*m_l\nu} {2\sqrt{2}sin^2\theta_wF_{\pi}m_{h_t^0}^2
}C(x_t)+\frac{V_{ts}m_lm_t^*m_b^{*2}M_W^2 }{4\sqrt{2}\nu g_2^4
F_{\pi}^3m_{h_t^0}^2 }C(x_s)\right). \label{rs}
\end{equation}
Here $x_s={m^*_t}^2/M_S^2$, $M_S$ is the mass of the top-pions and
$g_2$ is the $SU(2)$ coupling constant. $C(x_t)$ is the Inami-Lim
function in the $SM$ \cite{smf}. Since the neutral top-Higgs
coupling with fermions is different from that of neutral top-pion by
only a factor of $\gamma_5$, the expression of $R_P$ is same as that
of $R_S$ except only for the masses of the scalar particles. In our
numerical estimation, we will take $m_{\pi_t^0}=m_{h_t^0}=M_S$. In
this case, $R_P=R_S$.

The charged top-pions $\pi^{\pm}$ give contributions to the $SM$
function $Y(x)$ via the diagrams $(b)$, $(c)$ and $(d)$ in
Fig.\ref{a2}, the expression of the function $Y^{TC}(x_s)$ can be
written as:
\begin{eqnarray}
Y^{TC}(x_s)&=&\frac{1}{4\sqrt{2}G_F
F_{\pi}^2}\left[-\frac{x_s^3}{8(1-x_s)}-\frac{x_s^3}{8(1-x_s)^2}lnx_s\right].
\label{ytc}
\end{eqnarray}

\noindent{ \bf C. Numerical results}

The branching ratios of the decay processes $B_s \to l^+l^-$ can be
written as\cite{lp}:

\begin{equation}
Br(\bqll) = a_s \left[
  \left| 2 m_l R_A  - \frac{m_{B_s}^2}{m_b + m_s} R_P\right|^2
  + \left( 1 - \frac{4 m_{l}^{2}}{m_{B_s}^2}\right)
  \left| \frac{m_{B_s}^2}{m_b + m_s} R_S\right|^2
   \right]\;,
  \label{blep_gen}
\end{equation}
where
\begin{equation}
a_s \equiv \frac{G_F^2 \alpha^2}{64 \pi^3} \,
  \left | V_{ts}^\ast V_{tb}  \, \right |^2 \tau_{B_s} f_{B_s}^2 m_{B_s} \,
 \sqrt{ 1 - \frac{4 m_l^2 }{m_{B_s}^2} } \;.
\end{equation}
Here $\tau_{B_s}$ is the lifetime of $B_s$.

The longitudinal polarization asymmetry of the final leptons in $B_s
\to l^+ l^-$ is defined as follows \cite{lp}:

\begin{equation}
  A_{LP}^\pm \equiv
  \frac{
    \left[ \Gamma(s_{l^-},s_{l^+}) + \Gamma(\mp s_{l^-},\pm s_{l^+}) \right] -
    \left[ \Gamma(\pm s_{l^-},\mp s_{l^+}) + \Gamma(-s_{l^-},-s_{l^+})
    \right]}
    {\left[ \Gamma(s_{l^-},s_{l^+}) + \Gamma(\mp s_{l^-},\pm s_{l^+}) \right] +
    \left[ \Gamma(\pm s_{l^-},\mp s_{l^+}) + \Gamma(-s_{l^-},-s_{l^+}) \right]  }
    \;  ,
\end{equation}
$s_{l^{\pm}}$ are defined into one direction in dilepton rest frame
as $(0,\pm\frac{p_-}{|p_-|})$. For only one direction, there are no
difference between the final leptons, thus there is $A_{LP}^+ =
A_{LP}^- \equiv A_{LP}$. Then the $A_{LP}$ can be written as:
\begin{equation}
A_{LP}(B_s \to l^+l^-)  = \frac{ 2
              \sqrt{ 1- \frac{4 m_l^2 }{m_{B_s}^2} }
              Re \left[  \frac{m_{B_s}^2}{m_b + m_s} R_S \left( 2 m_l
                          R_A  - \frac{m_{B_s}^2}{m_b + m_s}~R_P \right)
                \right]  } {
          \left| 2 m_l R_A  - \frac{m_{B_s}^2}{m_b + m_s}  R_P \right|^2
      + (1-\frac{4m_l^2}{m_{B_s}^2})
            \left|  \frac{m_{B_s}^2}{m_b + m_s} R_S \right|^2 }.
  \label{eqn:alp}
  \end{equation}
$A_{LP}^{SM}(B_s \to l^+l^-) \simeq 0$ because
$R_S\sim\ord(10^{-5})$ in the $SM$.

\begin{table}
\begin{center}
\begin{displaymath}
\begin{tabular}{|l|l|}
\hline \spp $G_F = 1.166 \times 10^{-5} \; \gev^{-2}$ &
     $m_{B_s}=5.366 \; \gev$ \\
\spp $\alpha = 7.297 \times 10^{-3}$ &
     $ m_B=5.279 \; \gev$  \\
\spp $\tau_{B_s} = (1.437_{-0.030}^{+0.031}) \times 10^{-12} s$  &
     $V_{tb}= 1.0 $  \\
\spp $\tau_{B_d} = 1.53 \times 10^{-12} s$  &
     $V_{ts}= (40.6 \pm 2.7) \times 10^{-3}$ \\
\spp $m_{\mu}=0.105 \;\gev$ &
     $f_{B_s}=(0.259 \pm 0.027) \; \gev$ \cite{mackenzie}\\
\spp $M_W= 80.425(38) \; \gev$ & $sin^2\theta_w=0.23120(15)$
     \\ \hline
\end{tabular}
\end{displaymath}
\caption{Numerical inputs used in our
  analysis. Unless explicitly specified, they are \hspace*{1.7cm}taken from the
  Particle Data Group \cite{15}.\label{tab:inputs}}
\end{center}
\end{table}

Before giving numerical results, we need to specify the relevant
$SM$ parameters. These parameters have mainly been shown in Table
\ref{tab:inputs}. We take the coupling constant $K_1$, the model
dependent parameter $\varepsilon$, the mass of non-universal gauge
boson $M_{Z'}$ and the mass of scalars $M_{S}$ as free parameters in
our numerical estimation. The value of $M_{S}$ remains subject to
large uncertainty \cite{tc203}. However, it has been shown that its
value is generally allowed to be in the range of a few hundred $\rm
GeV$ depending on the models \cite{tc19}. In our numerical
estimation, we will assume that $M_S$ is in the range of $200\rm
GeV\sim\rm 500GeV$. The lower bounds on $M_{Z'}$ can be obtained
from dijet and dilepton production in the Tevatron experiments
\cite{25} or $B\bar{B}$ mixing \cite{26}. However, these bounds are
significantly weaker than those from the precision electroweak data.
Reference \cite{116} has shown that, to fit the precision
electroweak data, the $Z'$ mass $M_{Z'}$ must be larger than $1\
TeV$. In our numerical estimation, we will assume that the values of
the free parameters $\varepsilon$, $K_1$ and $M_{Z'}$ are in the
range of $0.03\sim 0.1$, $0\sim1$ and $1000\ \rm GeV\ \sim 2000\ \rm
GeV$, respectively.

\begin{figure}
\centering \subfigure[$K_1=0.4$]{
\includegraphics[scale=0.88]{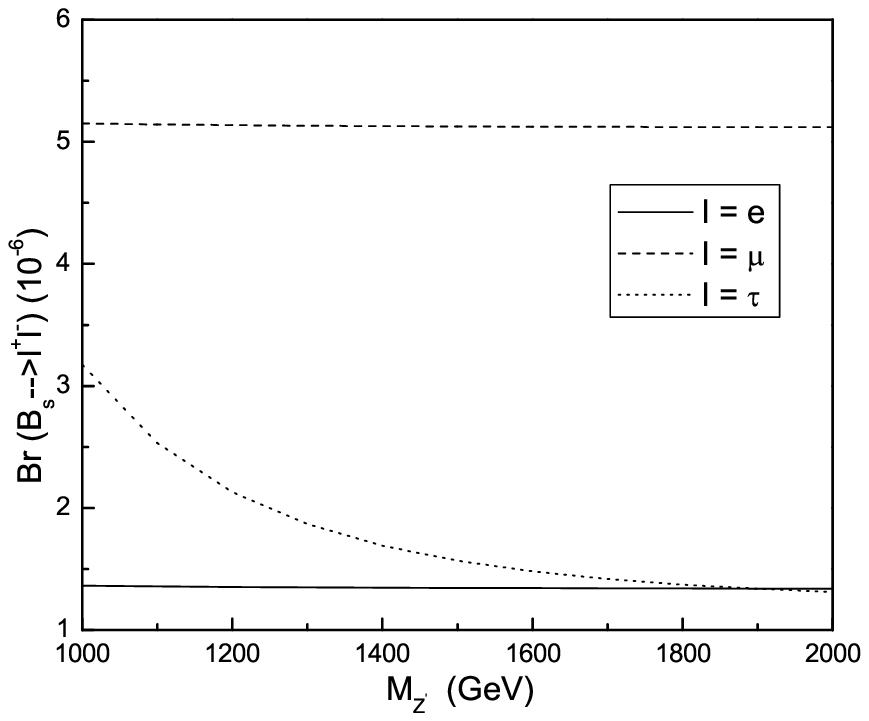}}
\subfigure[$K_1=0.8$]{
\includegraphics[scale=0.88]{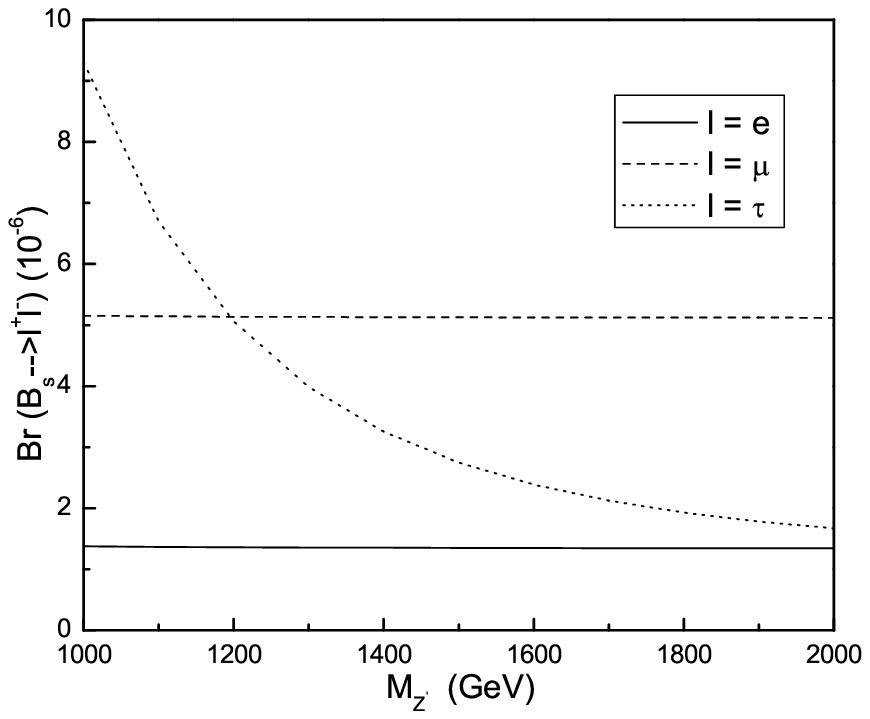}}
\caption{ The branching ratios of $B_s \to l^+l^-$ as function of
the parameter $M_{Z'}$ \hspace*{1.8cm}for $K_1=0.4$ (a) and
$K_1=0.8$ (b).} \label{fig:1}
\end{figure}

First we give our numerical results of the decay processes $B_s \to
l^+l^-$ induced by the non-universal gauge boson $Z'$. The branching
ratios of $B_s \to l^+l^-$ are plotted in Fig.\ref{fig:1} as
function of the mass parameter $M_{Z'}$ for $K_1=0.4$ and $0.8$, in
which we have multiplied the factors $10^{7}$ and $10^{3}$ to the
values of $Br(B_s \to e^+e^-)$ and $Br(B_s \to \mu^+\mu^-)$,
respectively. From these figures one can see that the values of
$Br(B_s \to \tau^+\tau^-)$ are sensitive to the mass of $Z'$, they
increase as the mass parameter $M_{Z'}$ decreasing. For $l=e,\mu$,
the values of their branching ratios are not so sensitive to the
parameter $M_{Z'}$. Because the contributions of $Z'$ to $Br(B_s \to
e^+e^-)$ and $Br(B_s \to \mu^+\mu^-)$ are small relative to the $SM$
contributions. The values of the corresponding branching ratios are
both below $\ord(10^{-9})$ which are not easy to be observed in
current collider experiments. The contributions of $Z'$ to the
branching ratio of the decay $B_s \to \tau^+\tau^-$ are large, since
the non-universal gauge boson $Z'$ has large couplings to the third
generation fermion with respect to the first two generations, it can
make the branching ratio value reach $\ord(10^{-6})$ with reasonable
values of the free parameters.

\begin{figure}
\centering \subfigure[$\varepsilon=0.04$]{
\includegraphics[scale=0.88]{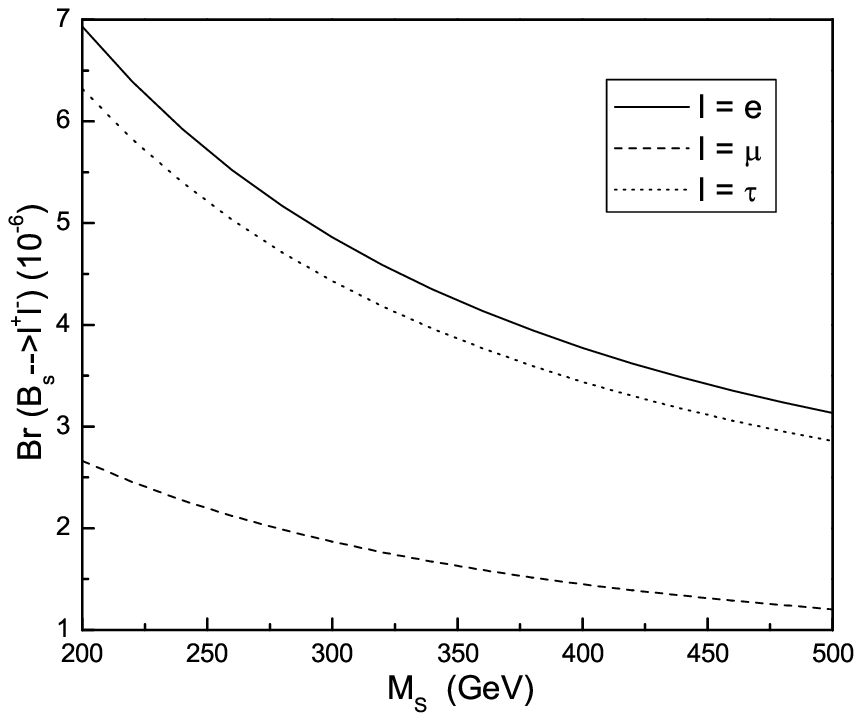}}
\subfigure[$\varepsilon=0.08$]{
\includegraphics[scale=0.88]{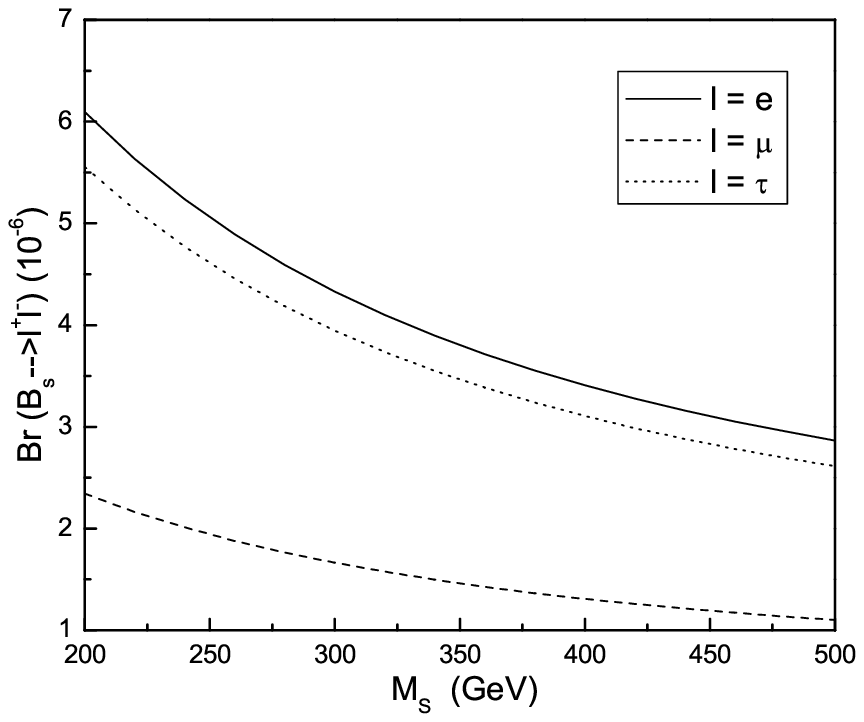}}
\caption{The branching ratios of $B_s \to l^+l^-$ as function of the
parameter $M_{S}$ for \hspace*{1.8cm}$\varepsilon=0.04$ (a) and
$\varepsilon=0.08$ (b).}\label{fig:2}
\end{figure}

\begin{figure}
\centering \subfigure[$\varepsilon=0.04$]{
\includegraphics[scale=0.85]{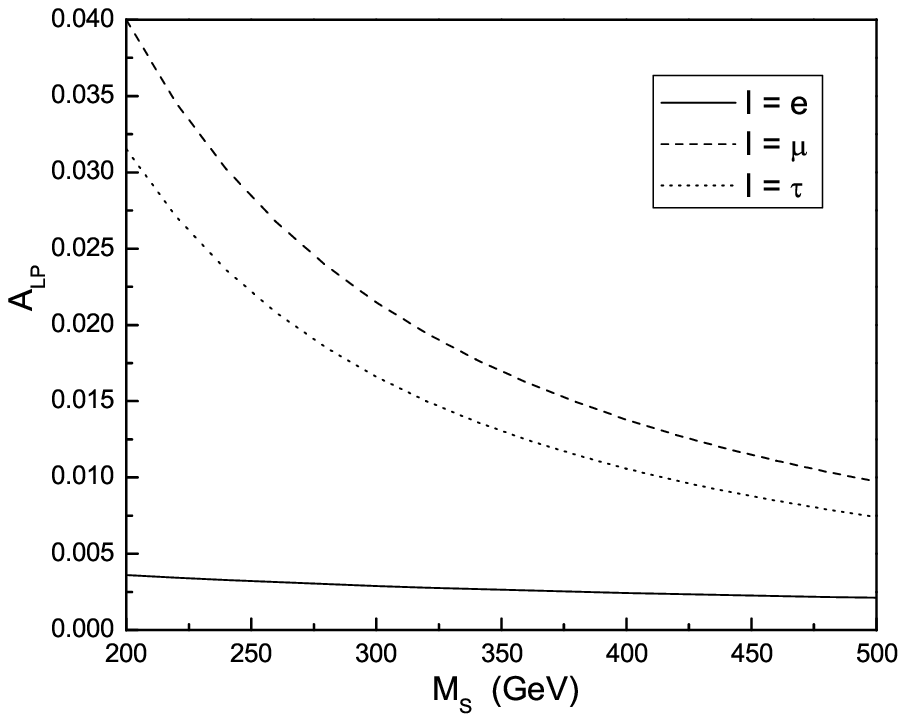}}
\subfigure[$\varepsilon=0.08$]{
\includegraphics[scale=0.85]{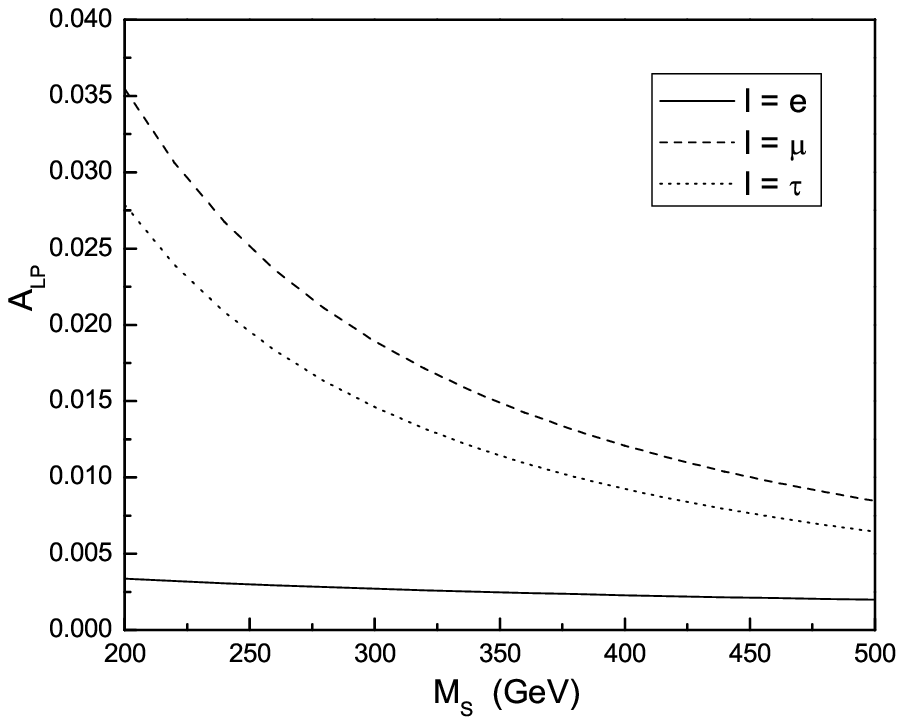}}
\caption{ The longitudinal polarization asymmetry in $B_s \to
l^+l^-$ as function of the \hspace*{1.8cm}parameter $M_S$ for
$\varepsilon=0.04$ (a) and $\varepsilon=0.08$ (b). } \label{fig:3}
\end{figure}

The branching ratios of $B_s \to l^+l^-$ contributed by the scalars
($\pi_t^{0,\pm}$ and $h_t^0$) are plotted in Fig.\ref{fig:2} as
function of the mass parameter $M_{S}$ for $\varepsilon=0.04$ and
$0.08$, in which we have multiplied the factors $10^{7}$ and
$10^{2}$ to the values of $Br(B_s \to e^+e^-)$ and $Br(B_s \to
\mu^+\mu^-)$, respectively. It is obvious that the values of the
branching ratios for these decays increase as the parameter $M_{S}$
decreasing. Furthermore, the enhancement to the branching ratio of
the decay process $B_s \to \mu^+\mu^-$ is larger than that of the
$Z'$ contributions by an order of magnitude.

The value of $Br(B_s \to e^+e^-)$ is smaller than that of $Br(B_s
\to \mu^+\mu^-)$ by five orders of magnitude, which is because it is
suppressed by $m_e^2/m_{\mu}^2$ with respect to $\mu$ channel. The
branching ratio for $\tau^+\tau^-$ mode is enhanced by a factor of
$10^2$ to $\mu$ channel, its value can reach $\ord(10^{-6})$ by our
calculation. However, the $\tau^+\tau^-$ channel is still not easy
to be observed under present experimental precision, while the
current experimental upper limit for $Br(B_s \to \tau^+\tau^-)$ from
the BARBAR collaboration is $4.1\times10^{-3}$ at $90\%~ \rm{C.L.}$
\cite{567}. So the experimental searches for $B_s \to l^+l^-$ have
focused on the $\mu$ channel, and we only discuss this channel.
Comparing with the $SM$ prediction $Br
(B_s\to\mu^+\mu^-)=3.86\pm0.15\times10^{-9}$ \cite{bdk}, the
contributions of the new scalars predicted by the $TC2$ model can
enhance this value by one order of magnitude, so our results are
more approach to the experimental data given by Eq.~(\ref{exp111}).

Obviously, the non-universal gauge boson $Z'$ has no contributions
to the $SPNP$ operators, so it was not considered in this
subsection. The longitudinal polarization asymmetry $A_{LP}$
contributed by the new scalars predicted by the $TC2$ model as
function of the parameter $M_S$ are plotted in Fig.\ref{fig:3}. From
these figures one can see that the $A_{LP}$ is sensitive to the mass
of the scalars, especially for $l=\mu,\tau$, however it is less
sensitive to the parameter $\varepsilon$. The values of the
asymmetry $A_{LP}$ can reach nearly $4\%$ for $l=\mu,\tau$ when the
mass of the scalars get to $200 \rm~ GeV$.

\noindent{\bf \large IV. The contributions of the $\rm{TC2}$ model
to the rare decay processes \hspace*{0.7cm}$B \to K l^+l^-$}

The effective Hamilton for the decay $B\to Kl^+l^-$ is similar to
that of $B_s \to l^+l^-$ as shown in Eq.~(\ref{lag-tot}), which is
constituted by two parts. The $SPNP$ part is same as the expression
shown in Eq.~(\ref{hsp}). In the framework of the $TC2$ model, The
$H_0$ part can be written as \cite{0805}:
\begin{equation}
\begin{split}
H_{0} &= \frac{\alpha G_F}{\sqrt{2} \pi} V_{tb} V^*_{ts} \biggl\{
C^{\rm eff}_9        (\bar{s} \gamma_\mu P_L b)    \, \bar{l}
\gamma_\mu l  +
C_{10}              (\bar{s} \gamma_\mu P_L b)        \,   \bar{l} \gamma_\mu \gamma_5 l \\
&- 2 \frac{C^{\rm eff}_7}{q^2} m_b \, (\bar{s} i \sigma_{\mu\nu}
q^\nu P_R b) \, \bar{l} \gamma_\mu l \biggr\} \, .
\end{split}
\label{SML}
\end{equation}
Here $q_\mu$ is the sum of $4$-momenta of $l^+$ and $l^-$.  The
Wilson coefficients $C_7^{eff}$, $C_9^{eff}$ and $C_{10}$ contain
two parts of contributions from the $SM$ and the $TC2$ model.

Similar to the decay processes $B_s \to l^+l^-$, the non-universal
gauge boson $Z'$ give contributions to the Wilson coefficients
$C_9^{eff}$ and $C_{10}$, the relevant Feynman diagrams are same as
Fig.\ref{a1} and the relevant functions $Y^{TC}(x_t)$ and
$Z^{TC}(x_t)$ have same expressions as shown in Eq.~(\ref{ytcz}).

The charged top-pions $\pi_t^{\pm}$ can give contributions to the
Wilson coefficients $C_7^{eff}$ and $C_9^{eff}$. The relevant
Feynman diagrams are similar to Fig.\ref{a2}. However, these penguin
diagrams are induced by $\gamma$ penguins, gluon penguins and
chromomagnetic penguins. The coefficients $C_7^{eff}$ and
$C_9^{eff}$ can be expressed in terms of the corresponding functions
$D_1(x_s)$, $E_1(x_s)$ and $E_1'(x_s)$, which are added to the
corresponding $SM$ functions $D_0(x_t)$, $E_0(x_t)$ and $E_0'(x_t)$
\cite{smf}. The detailed expression forms of the these functions are
\cite{tc301}:
\begin{eqnarray}
D_1(x)&=&\frac{1}{4\sqrt{2}G_FF_{\pi}}\left(\frac{47-79x+38x^2}{108(1-x)^3}+\frac{3-6x^2+4x^3}
{18(1-x)^4}ln(  x)\right),\\
E_1(x)&=&\frac{1}{4\sqrt{2}G_FF_{\pi}}\left(\frac{7-29x+16x^2}{36(1-x)^3}-\frac{3x^2-2x^3}{6(1-x)^4}ln  (x)\right),\\
E_1'(x)&=&\frac{1}{8\sqrt{2}G_FF_{\pi}}\left(\frac{5-19x+20x^2}{6(1-x)^3}-\frac{x^2-2x^3}{(1-x)^4}ln(
x)\right).
\end{eqnarray}
We can obtain the corrected Wilson coefficients $C_7^{eff}$,
$C_9^{eff}$ and $C_{10}$ with these corrected functions using the
relevant expressions of these coefficients in References
\cite{ued,tc301}, which are listed in Appendix C. The neutral
top-pion $\pi_t^0$ and top-Higgs $h_t^0$ can also give contributions
to these decay processes through the $SPNP$ operators, and the
expression forms of $R_S$ ($R_P$) are same as those shown in
Eq.~(\ref{rs}).

\begin{figure}
\centering \subfigure[$K_1=0.4$]{
\includegraphics[scale=0.88]{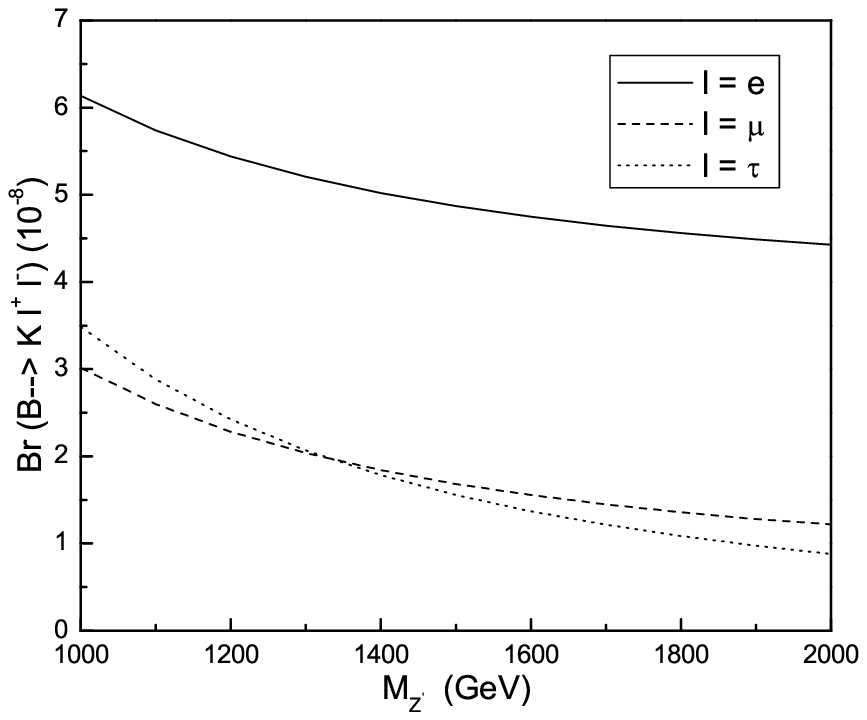}}
\subfigure[$K_1=0.8$]{
\includegraphics[scale=0.88]{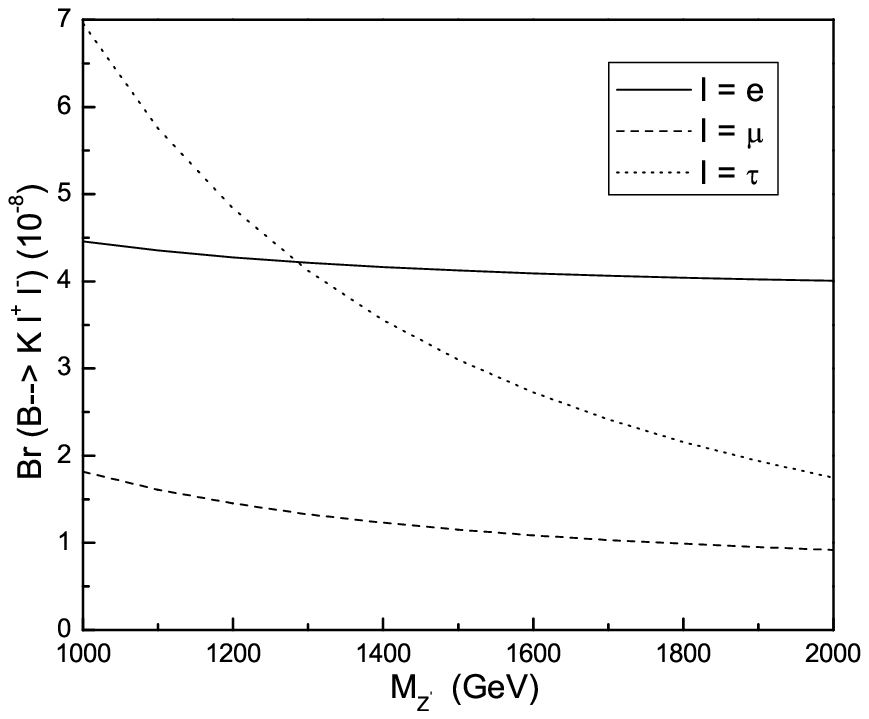}}
\caption{ The branching ratios of $B \to K l^+l^-$ as function of
the parameter $M_{Z'}$ \hspace*{1.8cm}for $K_1=0.4$ (a) and
$K_1=0.8$ (b). } \label{fig:4}
\end{figure}

The branching ratios $Br(B \to K l^+l^-)$ ($l=e,\mu$ and $\tau$)
contributed by the gauge boson $Z'$ are plotted in Fig.\ref{fig:4}
as a function of the mass parameter $M_{Z'}$ for two values of
$K_1$, in which we have multiplied the factor $10^{-1}$ and
$10^{-2}$ to the branching ratios of decays $B \to K \mu^+\mu^-$ and
$B \to K \tau^+\tau^-$ respectively. From this figure one can see
that the values of the branching ratios for $l=e,\mu$ and $\tau$
increase as the parameter $M_{Z'}$ decreasing. However, the
branching ratios for $l=e$ are not sensitive to the parameter
$M_{Z'}$ as shown in these figures. The values of the branching
ratios for $l=e$ and $\mu$ are not sensitive to the parameter $K_1$.
For $K_1=0.4$ and $1000\rm GeV\leq M_{Z'}\leq 2000\rm GeV$, the
values of $Br(B \to K e^+e^-)$ and $Br(B \to K \mu^+\mu^-)$ are in
the range of $6.1\times10^{-8}\sim4.4\times10^{-8}$ and
$3.0\times10^{-7}\sim1.2\times10^{-7}$, respectively.

\begin{figure}
\centering \subfigure[$\varepsilon=0.04$]{
\includegraphics[scale=0.88]{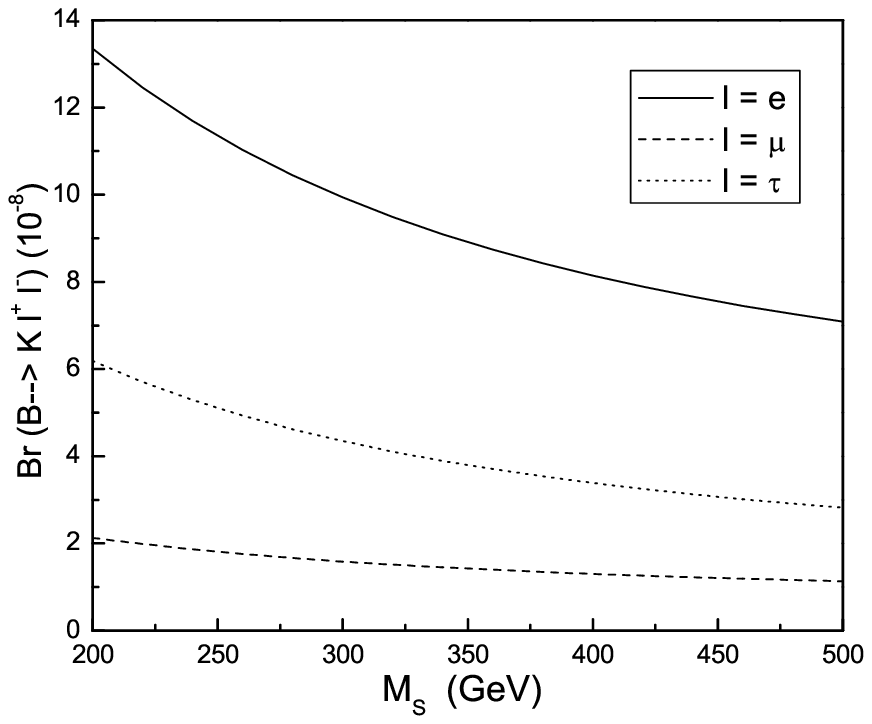}}
\subfigure[$\varepsilon=0.08$]{
\includegraphics[scale=0.88]{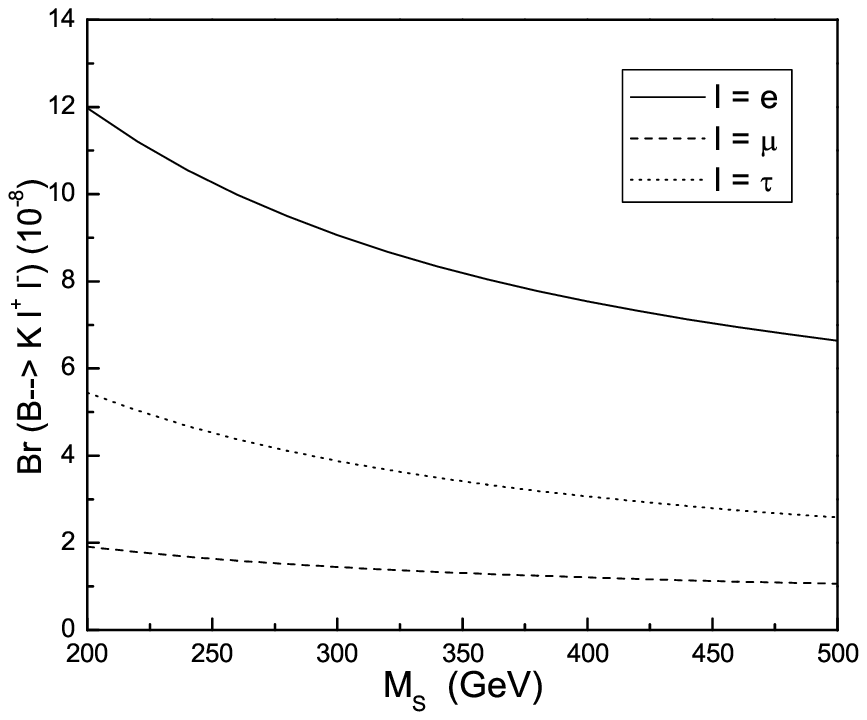}}
\caption{ The branching ratios of $B \to K l^+l^-$ as function of
the parameter $M_S$ for \hspace*{1.8cm}$\varepsilon=0.04$ (a) and
$\varepsilon=0.08$ (b). } \label{fig:8}
\end{figure}

The branching ratios of the decay processes $B \to K l^+l^-$
contributed by the scalars ($\pi_t^{0,\pm}, h_t^0$) are plotted in
Fig.\ref{fig:8} as function of the mass parameter $M_{S}$ for
$\varepsilon=0.04$ and $0.08$, in which we have multiplied the
factors $10^{-1}$ to the branching ratio of $B \to K \mu^+\mu^-$.
From these figures, one can see that the values of the branching
ratios of these decay processes increase as the parameter $M_{S}$
decreasing. All of their values are not sensitive to the parameter
$\varepsilon$. The contributions of the scalars for $l=e$ and $\mu$
are comparable to those of the non-universal gauge boson $Z'$, the
values of the branching ratios of $B \to K e^+e^-$ and $B \to K
\mu^+\mu^-$ contributed by both the scalars and the non-universal
gauge boson can reach $\ord(10^{-7})$, which give an explanation to
the deviation between the experimental data and the $SM$ predictions
in Reference \cite{lunghi}. While the scalar's contribution to the
decay process $B \to K \tau^+\tau^-$ is smaller than that of the
non-universal gauge boson $Z'$ by two order of magnitude and
therefore can be neglected. When the $Z'$ mass is in the range of
$1000\rm~ GeV\sim2000\rm~ GeV$, the values of $Br(B \to K
\tau^+\tau^-)$ are in the range of
$7.0\times10^{-6}\sim1.7\times10^{-6}$. This result is 2 orders of
magnitude larger than the $e$ and $\mu$ channel, which is because of
the large coupling of $Z'$ to the third generation fermions.


The normalized forward-backward ($FB$) asymmetry can be defined as
\cite{fb}:
\begin{eqnarray}
A_{FB}(z)= \frac{\int_0^{1}dcos\theta \frac{d^2\Gamma}{dz
dcos\theta}-\int_{-1}^{0}dcos\theta \frac{d^2\Gamma}{dz dcos\theta}}
{\int_0^{1}dcos\theta \frac{d^2\Gamma}{dz
dcos\theta}+\int_{-1}^{0}dcos\theta \frac{d^2\Gamma}{dz
d\cos\theta}}\;.
\end{eqnarray}
After the integral calculation of $FB$ asymmetry gives,
\begin{equation}
\langle A_{FB} \rangle = \frac{2 \tau_B \Gamma_{0} \, \hat{m}_{l} \,
\beta_{\mu}^{2} \, R_S \, \int dz \, a_{1}(z) \, \phi(1,k^2,z)}{Br(B
\to K l^+ l^-)} \; , \label{avg-afb}
\end{equation}
where $\tau_B$ is the lifetime of $B$ meson and $Br(B \to K l^{+}
l^{-})$ is the total branching ratio of $B \to K l^{+} l^{-}$ and
$\Gamma_{0}$ is the total width of the $B$ meson, which can be
written as:
\begin{equation}
\Gamma_{0}=\frac{G_{F}^{2}\alpha^{2}}{2^{9}\pi^{5}}\,|V_{tb}V_{ts}^{*}|^{2}\,
m_{B}^{5} \;,
\end{equation}
\begin{equation}
a_{1}(z)=\frac{1}{2}(1-k^{2})C_{9}f_{0}(z)f_{+}(z)+(1-k)C_{7}f_{0}(z)f_{T}(z)\;.
\end{equation}
Other relevant functions such as $\phi(1,k^2,z)$ are listed in
Appendix C. The form factors $f_{+}$, $f_0$ and $f_T$ are defined in
the relevant matrix elements as:
\begin{eqnarray}
\left< K(p') \left|\bar{s}\gamma_{\mu}b\right|
B(p)\right>&=&(2p-q)_{\mu}f_{+}(z)+(\frac{1-k^2}{z})\,
q_{\mu}[f_{0}(z)-f_{+}(z)]\;, \\
\left< K(p')\left|\bar{s}i\sigma_{\mu\nu}q^{\nu}b\right|
B(p)\right>&=&-\Big[(2p-q)_{\mu}q^2-(m_{B}^{2}-m_{K}^{2})q_{\mu}\Big]\,\frac{f_{T}(z)}{m_B+m_{K}}\;,\\
\left< K(p')\left|\bar{s}b\right|
B(p)\right>&=&\,\frac{m_B(1-k^2)}{\hat{m}_b}\,f_{0}(z)\; .
\end{eqnarray}
Here, $k \equiv m_K/m_B$, $z \equiv q^2/m_{B}^{2}$ and $\hat{m}_b
\equiv m_b/m_B$. The form factors $f_{+}$, $f_0$ and $f_T$ can be
calculated by using the light cone $QCD$ approach. Their particular
forms can be found in Reference \cite{0805}. In this paper, we
assume $\hat{m}_b\doteq1$.

\begin{figure}
\centering \subfigure[$\varepsilon=0.04$]{
\includegraphics[scale=0.88]{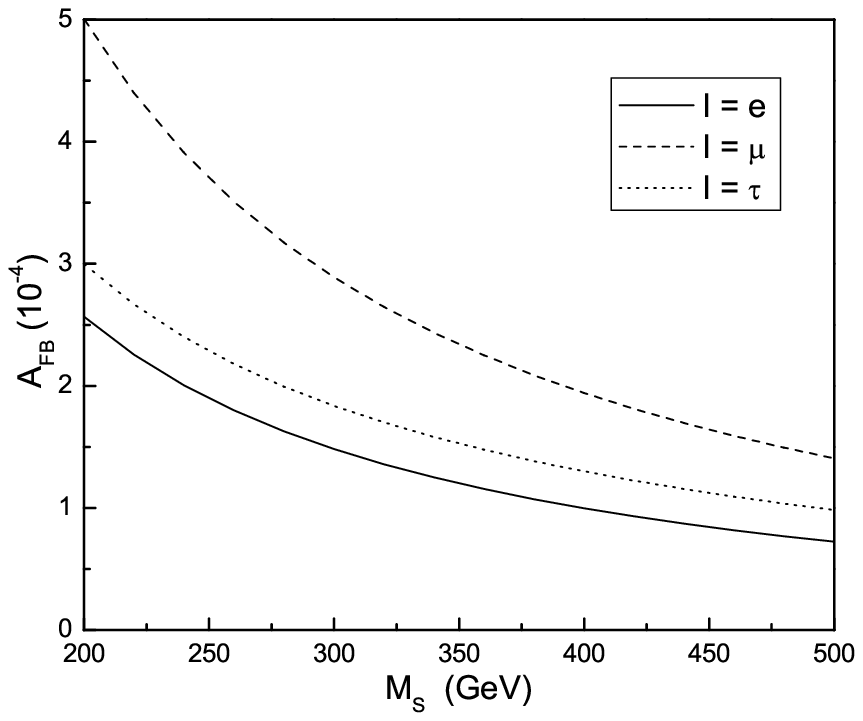}}
\subfigure[$\varepsilon=0.08$]{
\includegraphics[scale=0.88]{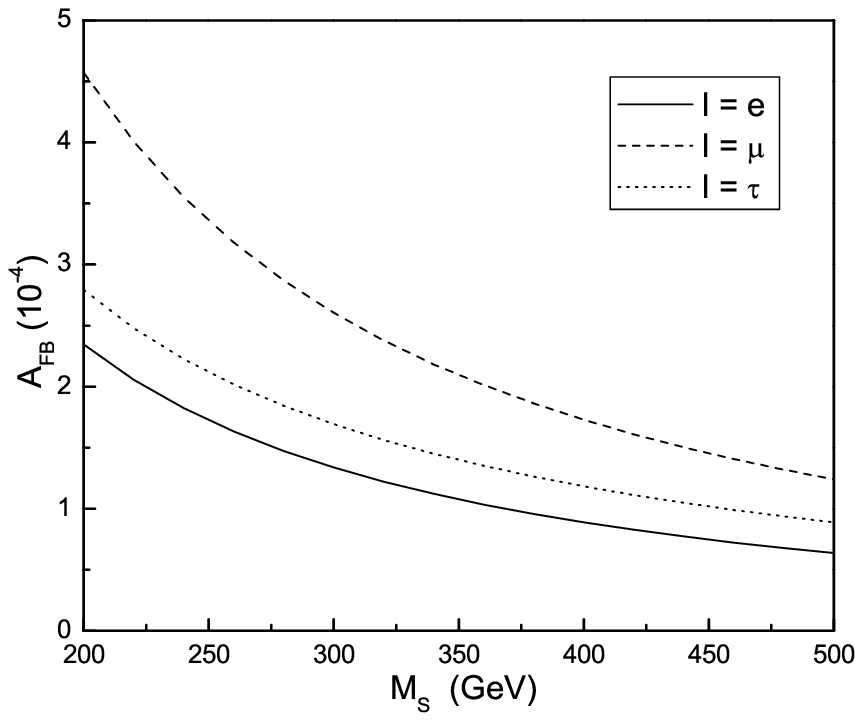}}
\caption{ In the $TC2$ model, the forward-backward asymmetry in $B
\to Kl^+l^-$ as \hspace*{1.8cm}function of $M_S$ for the parameter
$\varepsilon=0.04$ (a) and $\varepsilon=0.08$ (b).} \label{fig:9}
\end{figure}

The production of the $FB$ asymmetries are only sensitive to $SPNP$
operators. From Eq.~(\ref{avg-afb}), one can see that the
non-universal gauge boson $Z'$ has no contribution to the $FB$
asymmetry, so we only discuss the contributions coming from the
scalars ($\pi_t^{0,\pm}, h_t^0$).

The $FB$ asymmetry $A_{FB}$ of leptons in the decay processes $B\to
Kl^+l^-$ are plotted in Fig.\ref{fig:9} as function of the parameter
$M_S$ for $\varepsilon=0.04$ and $0.08$, in which we have multiplied
the factors $10^{5}$ and $10$ to the value of $A_{FB}(B \to K
e^+e^-)$ and $A_{FB}(B \to K \mu^+\mu^-)$ respectively. From this
figure one can see that the value of $A_{FB}$ is smaller than
$\ord(10^{-3})$ in most of the parameter spaces. Comparing its
experimental measurement value, this value is not large enough to be
observed in experiments. One can see that the contributions of the
$TC2$ model to the $FB$ asymmetry in these decay processes are
smaller than those of the $SUSY$ models. Considering the uncertainty
in measurements, it is very difficult to detect the signals of the
$TC2$ model through measuring the $FB$ asymmetry about these decay
processes.

\noindent{\bf \large V. The contributions of the $\rm{LHT}$ model to
the rare decay processes \hspace*{0.7cm}$b \to s l^+l^-$}

The $LHT$ model \cite{6} is based on an $SU(5)/SO(5)$ global
symmetry breaking pattern. A subgroup $[SU(2)\times U(1)]_{1}\times
[SU(2)\times U(1)]_{2}$ of the $SU(5)$ global symmetry is gauged,
and at the scale $f$ it is broken into the $SM$ electroweak symmetry
$SU(2)_{L}\times U(1)_{Y}$. T-parity is an automorphism which
exchanges the $[SU(2)\times U(1)]_{1}$ and $[SU(2)\times U(1)]_{2}$
gauge symmetries. The T-even combinations of the gauge fields are
the $SM$ electroweak gauge bosons $W_{\mu}^{a}$ and $B_{\mu}$. The
T-odd combinations are T-parity partners of the $SM$ electroweak
gauge bosons.

After taking into account $EWSB$, at the order of $\nu^{2}/f^{2}$,
the masses of the T-odd set of the $SU(2)\times U(1)$ gauge bosons
are given as:
\begin{equation}
M_{B_{H}}=\frac{g'f}{\sqrt{5}} [1-\frac{5\nu^{2}}{8f^{2}}],
\hspace{0.5cm}M_{Z_{H}} \approx
M_{W_{H}}=gf[1-\frac{\nu^{2}}{8f^{2}}],
\end{equation}
where $f$ is the scale parameter of the gauge symmetry breaking of
the $LHT$ model. $g'$ is the $SM$ $U(1)_{Y}$ gauge coupling
constants. Because of the smallness of $g'$, the T-odd gauge boson
$B_{H}$ is the lightest T-odd particle, which can be seen as an
attractive dark matter candidate \cite{16}. To avoid severe
constraints and simultaneously implement T-parity, it is necessary
to double the $SM$ fermion doublet spectrum \cite{6,17}. The T-even
combination is associated with the $SU(2)_{L}$ doublet, while the
T-odd combination is its T-parity partner. The masses of the T-odd
fermions can be written in a unified manner as:
\begin{equation}
M_{F_{i}}=\sqrt{2}k_{i}f,
\end{equation}
where $k_{i}$ are the eigenvalues of the mass matrix $k$ and their
values are generally dependent on the fermion species $i$.

The mirror fermions (T-odd quarks and T-odd leptons) have new flavor
violating interactions with the $SM$ fermions mediated by the new
gauge bosons $(B_{H},W_{H}^{\pm}$, or $Z_{H})$, which are
parametrized by four $CKM$-like unitary mixing matrices, two for
mirror quarks and two for mirror leptons \cite{11,12}:
\begin{equation}
V_{Hu},\hspace*{0.2cm}V_{Hd},\hspace*{0.2cm}V_{Hl},\hspace*{0.2cm}V_{H\nu}.
\end{equation}
They satisfy:
\begin{equation}
V_{Hu}^{+}V_{Hd}=V_{CKM},\hspace*{0.2cm}V_{H\nu}^{+}V_{Hl}=V_{PMNS}.
\end{equation}
Where the $CKM$ matrix $V_{CKM} $ is defined through flavor mixing
in the down-type quark sector, while the $PMNS$ matrix $V_{PMNS} $
is defined through neutrino mixing.


The contributions of the $LHT$ model to the rare decay processes $b
\to s l^+l^-$ are mainly coming from the corrections to the Wilson
coefficients, which related to the $SM$ Inami-Lim functions
\cite{smf}. The branching ratios of the decay processes $B_s \to
l^+l^-$ in the $SM$ depend on a function $Y_{SM}$ and the $LHT$
effects enter through the modification of the function $Y_{SM}$
\cite{11}. With the $LHT$ effects $Y_{SM}$ is replaced by
\cite{lht}:
\begin{equation}
Y_s=Y_{SM} + \bar Y^{\rm even} +\frac{\bar Y_s^{\rm
odd}}{\lambda_t^{(s)}},
\end{equation}
where $\bar Y^{\rm even}$ and $\bar Y_s^{\rm odd}$ represent the
effects from T-even and T-odd particles, respectively. The branching
ratios normalized to the $SM$ predictions are then given by:
\begin{equation}
\frac{Br(B_s\to l^+l^-)}{Br(B_s\to l^+l^-)_\text{SM}} =
\left|\frac{Y_s}{Y_\text{SM}}\right|^2,
\end{equation}
which $Br(B_s\to l^+l^-)_\text{SM}$ are the  branching ratios
predicted by the $SM$. Their particular numerical values of the
branching ratios for the decay processes $B_s\to l^+l^-$ in the
$LHT$ model are listed as follows:
\begin{eqnarray}
Br(B_s\to e^+e^-)&=&(1.36\pm0.05)\times10^{-13},\\
Br(B_s\to \mu^+\mu^-)&=&(5.79\pm0.23)\times10^{-9},\\
Br(B_s\to \tau^+\tau^-)&=&(1.23\pm0.05)\times10^{-6}.
\end{eqnarray}

The branching ratios of the decay processes $B \to K l^+l^-$ in the
$SM$ depend on the functions $Y_{SM}$, $Z_{SM}$ and $D_0'(x_t)$
($D_0'(x_t)$ is same as in $B \to X_s \gamma$ \cite{lht}), the $LHT$
effects enter through the modification of these functions. The
modifications of the function $Y_{SM}$ has been given above, and the
modifications of the function $Z_{SM}$ is given by \cite{lht,11}:
\begin{equation}
Z_s= Z_{SM} + \bar Z^{\rm even} + \frac{\bar
Z_s^{odd}}{\lambda_t^{(s)}},
\end{equation}
where $\bar Z^{\rm even}$ and $\bar Z_s^{odd}$ represent the effects
coming from T-even and T-odd particles, respectively. Similar with
Sec. IV, we can calculate the contributions of the $LHT$ model to
the decay processes $B \to K l^+l^-$. With reasonable values of the
free parameters in the framework of the $LHT$ model, the maximum
values of the branching ratios for the rare decays $B \to K l^+l^-$
are:
\begin{eqnarray}
Br(B\to K e^+e^-)&=&9.66\times10^{-6},\label{42}\\
Br(B\to K\mu^+\mu^-)&=&6.56\times10^{-6},\label{43}\\
Br(B\to K\tau^+\tau^-)&=&2.99\times10^{-7}.\label{44}
\end{eqnarray}

These numerical results are obtained by calculating the relative
correction to the $SM$ predictions in the framework of the $LHT$
model, while the $SM$ predictions exist the uncertainty coming from
the next-to-leading logarithmic ($NLO$) contributions and the
long-distance contributions, for which the $Br(B \to K l^+l^-)$ are
a little disparity away from their respective experimental upper
limits \cite{1201}. However, there is no disagreement with
experiment in some parameter ranges while the corrected effects is
no more than 15 percent.

The contributions of the $LHT$ model to the asymmetry observables
$A_{FB}$ and $A_{LP}$ in the rare decay processes $b \to s l^+l^-$
mainly come from the new neutral scalar particles. For the $B_s$
meson, there is an unitarity relation of the $V_{Hd}$ matrix
\cite{11}:
\begin{equation}
\xi^{(s)}_1+\xi^{(s)}_2+\xi^{(s)}_3=0 ,
\end{equation}
where $\xi^{(s)}_i=V_{Hd}^{*ib}V_{Hd}^{is}$. Considering this
relation, the calculations of the relevant Feynman diagrams similar
to Fig.\ref{a2} equal to zero. Hence, in the framework of the $LHT$
model, the total contributions induced by the neutral scalars equal
to zero. The contributions to the $A_{FB}$ and $A_{LP}$ is close to
the predictions in the $SM$.

\noindent{\bf \large VI. Conclusions }

The $SM$ is a very successful theory but it can only be an effective
theory below some high energy scales. To completely avoid the
problems arising from the elementary Higgs field in the $SM$,
various kinds of dynamical electroweak symmetry breaking models have
been proposed, among which the topcolor scenario is attractive
because it can explain the large top quark mass and provide a
possible $EWSB$ mechanism. The $TC2$ model has all essential
features of the topcolor scenario. It is expected that the possible
signals of the $TC2$ model should be detected in the future high
energy collider experiments.

In this paper we consider the contributions of the $TC2$ model to
observables related to the decay processes $B_s\to l^+l^-$ and $B\to
Kl^+l^-$. We find that the $TC2$ model can enhance the branching
ratios of the $SM$ predictions for these decay processes $B_s\to
l^+l^-$ and $B\to Kl^+l^-$. In wide ranges of the free parameter
space, it is possible to enhance the values of $Br(B_s\to l^+l^-)$
and $Br(B\to Kl^+l^-)$ by one order of magnitude. In the $TC2$
model, the non-universal gauge boson $Z'$ gives main contributions
to $Br(B_s\to \tau^+\tau^-)$, while the contributions of $Z'$ to
$Br(B_s\to e^+e^-)$ and $Br(B_s\to \mu^+\mu^-)$ are comparable with
those of the new scalars ($\pi_t^{0,\pm}, h_t^0$). For the decay
processes $B\to Ke^+e^-$ and $B\to K\mu^+\mu^-$, the contributions
of $Z'$ are comparable with those of the scalars. While the
contributions of the $TC2$ model to $Br(B\to K\tau^+\tau^-)$ mainly
come from $Z'$.

The production of the asymmetries are only sensitive to $SPNP$
operators, so there are no contributions of $Z'$ to the relevant
observables. We further calculate the contributions of the new
scalars predicted by the $TC2$ model to the asymmetry observables
$A_{FB}$ and $A_{LP}$ of leptons in the decay processes $B_s\to
l^+l^-$ and $B\to Kl^+l^-$. Our numerical results show that, when
the mass of the scalars gets to $200 \rm GeV$, the values of the
asymmetry $A_{LP}$ in the decay processes $B_s\to \mu^+\mu^-$ and
$B_s\to \tau^+\tau^-$ can reach $4\%$ . We hope that the values of
$A_{LP}$ for $l=\mu,\tau$ can approach the detectability threshold
of the near future experiments. However, the contributions of these
new scalars to $A_{FB}$ are around $\ord(10^{-4})$ in most of the
parameter space, which are not large enough to be detected.

The $LHT$ model is one of the attractive little Higgs models, which
satisfies the electroweak precision data in most of the parameter
space. This model can produce rich phenomenology at present and in
future high energy experiments. New particles predicted by this
model give contributions to the branching ratios of the rare decay
processes $B_s\to l^+l^-$ and $B\to Kl^+l^-$. Reference \cite{lht}
has shown that, comparing with their $SM$ predictions, the branching
ratios of the decay processes $B_s\to l^+l^-$ and $B\to Kl^+l^-$ can
be enhanced by at most $50\%$ and $15\%$, respectively. For
comparison, we give a brief description and particular numerical
results about these rare decays. In addition, we show that the
neutral scalars predicted by this model can not give contributions
to the asymmetry observables $A_{FB}$ and $A_{LP}$.

In conclusion, the effects of the $TC2$ model on the branching
ratios and asymmetry observables related to the rare decay processes
$b \to sl^+l^-$ can give positive contributions to the $SM$
predictions. The numerical results show that the branching ratios
for these decays are much close to the experimental data, such as
$Br(B_s\to \mu^+\mu^-)$. The value of $Br(B\to K\tau^+\tau^-)$ is
larger than the $SM$ prediction by one order of magnitude, which is
hoped to be observed in the future high accuracy experiments, or the
future experimental results may give constraints on the free
parameters of the $TC2$ model. Hence, it is indicated that the
possible signals of the $TC2$ model may be observed through the
above decay processes in future experiments.

\noindent{\bf \large Acknowledgments}

This work was supported in part by the National Natural Science
Foundation of China under Grants No.10675057, Specialized Research
Fund for the Doctoral Program of Higher Education(SRFDP)
(No.200801650002), the Natural Science Foundation of the Liaoning
Scientific Committee(No.20082148), and Foundation of Liaoning
Educational Committee(No.2007T086). \vspace{1.0cm}

\noindent{\bf \large Appendix}

\noindent{ \bf A. Relevant functions in the $SM$}

In this Appendix we list the functions in the $SM$ that entered the
present study of rare B decays.
\begin{eqnarray}
Y^{SM}(x)&=&\frac{1}{8}\left[\frac{x-4}{x-1}+\frac{3x}{(x-1)^2}\ln x\right],\\
Z^{SM}(x_t)&=&-\frac{1}{9}\ln x_t+\frac{18 x_t^4-163x_t^3+259
  x_t^2-108x_t}{144(x_t-1)^3}\nonumber\\
&&+\frac{32x_t^4-38x_t^3-15x_t^2+18x_t}{72(x_t-1)^4}\ln x_t\,,
\end{eqnarray}
\begin{eqnarray}
D_0(y)&=&-\frac{4}{9}\ln
y+\frac{-19y^3+25y^2}{36(y-1)^3}+\frac{y^2(5y^2-2y-6)}{18(y-1)^4}\ln
y\,,\\
E_0(y)&=&-\frac{2}{3}\ln y + \frac{y^2(15-16y+4y^2)}{6(y-1)^4}\ln y
+ \frac{y(18-11y-y^2)}{12(1-y)^3}\,,\\
 D'_0(y)&=&-\dfrac{(3y^3-2y^2)}{2(y-1)^4}\ln y +
\dfrac{(8y^3+5y^2-7y)}{12(y-1)^3}\,,\\
E'_0(y)&=&\dfrac{3y^2}{2(y-1)^4}\ln y +
\dfrac{(y^3-5y^2-2y)}{4(y-1)^3}\,.
\end{eqnarray}

\noindent{ \bf B. Relevant functions in the TC2 model}

In this Appendix we list the functions that entered the present
study of rare B decays in the framework of the TC2 model.
\begin{eqnarray}
C_{ab}(x)&=&-\frac{2g^2c_w^2F_1(x)}{3g_2^2(v_d+a_d)},\\
C_{c}(x)&=&\frac{2f^2c_w^2}{g_2^2}\left(\frac{2F_2(x)}{3(v_u+a_u)}+\frac{F_3(x)}{6(v_u-a_u)}\right),\\
C_{d}(x)&=&\frac{2f^2c_w^2}{g_2^2}\left(\frac{2F_4(x)}{3(v_u+a_u)}+\frac{F_5(x)}{6(v_u-a_u)}\right),\\
C(x)&=&\frac{F_1(x)}{-(0.5(Q-1)s_w^2+0.25)}.
\end{eqnarray}
Here the variables are defined as: $g=\sqrt{4\pi K_1}$,
$v_{u,d}=I_3-2Q_{u,d}s_w^2$, $s_w=\sin \theta _{w}$, $a_{u,d}=I_3$,
where $u, d$ represent the up and down type quarks, respectively.
\begin{eqnarray}
F_1(x)&=&-(0.5(Q-1)s_w^2+0.25)(x^2 ln(x)/(x-1)^2-x/(x-1)
-x(0.5(-0.5772\nonumber\\ &+&ln(4\pi)-ln(M_W^2))
+0.75-0.5(x^2ln(x)/(x-1)^2-1/(x-1)))),\\
F_2(x)&=&(0.5 Q s_w^2-0.25)(x^2ln(x)/(x-1)^2-2x
ln(x)/(x-1)^2+x/(x-1)),\\
F_3(x)&=&-Q s_w^2(x/(x-1)-x ln(x)/(x-1)^2),\\
F_4(x)&=&0.25(4 s_w^2/3-1)(x^2ln(x)/(x-1)^2-x-x/(x-1)),\\
F_5(x)&=&-0.25Q s_w^2x(-0.5772+ln(4\pi)-ln(M_W^2)+1-x
ln(x)/(x-1))\nonumber\\
&-&s_w^2/6(x^2ln(x)/(x-1)^2-x-x/(x-1)).
\end{eqnarray}

\noindent{ \bf C. Relevant expressions in our calculation}

In this Appendix we list the functions that entered the present
study of rare B decays and some expressions of the relevant
coefficients.
\begin{eqnarray}
M\,(B\rightarrow K l^{+}l^{-}) &=& \frac{\alpha G_F}{2\sqrt{2} \pi}
V_{tb} V^*_{ts}
\nonumber \\
&\times& \Bigg[\left< K(p')
\left|\bar{s}\gamma_{\mu}b\right|B(p)\right> \left\{C_{9}^{\rm
eff}\bar{u}(p_+)\gamma_{\mu}v(p_-)
+C_{10}\bar{u}(p_+)\gamma_{\mu}\gamma_{5} v(p_-)\right\}
\nonumber \\
& & - 2\frac{C^{\rm eff}_7}{q^2} m_b \left< K(p')\left|\bar{s}
i\sigma_{\mu\nu}q^{\nu}b\right|B(p)\right>
\bar{u}(p_+)\gamma_{\mu}v(p_-)
\nonumber \\
& & + \left< K(p')\left|\bar{s}b\right|B(p)\right>
\left\{R_S\bar{u}(p_+)v(p_-)+
R_P\bar{u}(p_+)\gamma_{5}v(p_-)\right\} \Bigg]\; ,
\end{eqnarray}
\begin{eqnarray}
\frac{d^{2}\Gamma}{dzdcos\theta} & = & \,\frac{G_{F}^{2}\alpha^{2}}{2^{9}\pi^{5}}\,|V_{tb}
V_{ts}^{*}|^{2}\, m_{B}^{5}\,\phi^{1/2}(1,k^2,z)\,\beta_{\mu}\nonumber \\
 & \times & \Bigg[\Big(\left|A\right|^{2}\beta_{\mu}^{2}+\left|B\right|^{2}\Big)z+\frac{1}{4}
 \phi(1,k^2,z)\Big(\left|C\right|^{2}+\left|D\right|^{2}\Big)(1-\beta_{\mu}^{2} \cos^{2}\theta)\nonumber \\
 &  & + 2\hat{m_{l}}(1-k^{2}+z)Re(BC^{*}) +
4\hat{m_{l}}^{2}\left|C\right|^{2}\nonumber \\
 &  &
+2\hat{m_{l}}\,\phi^{\frac{1}{2}}(1,k^2,z)\,\beta_{\mu}\,
Re(AD^{*})\, \cos\theta \Bigg] \;, \label{double_drate}
\end{eqnarray}
\begin{eqnarray}
A & \equiv & \frac{\,1}{2}(1-k^{2})f_{0}(z)R_S\;, \nonumber \\
B &\equiv
&-\hat{m_{l}}C_{10}\left\{f_{+}(z)-\frac{1-k^{2}}{z}(f_{0}(z)-f_{+}(z))\right\}+
\frac{1}{2}(1-k^{2})f_{0}(z)R_P\;, \nonumber \\
C &\equiv & C_{10}\,f_{+}(z)\;, \nonumber \\
D &\equiv & C_{9}^{eff}\, f_{+}(z)+2\, C_{7}^{eff}\,\frac{f_{T}(z)}{1+k}\;, \nonumber \\
\phi(1,k^2,z) & \equiv& 1+k^{4}+z^{2}-2(k^{2}+k^{2}z+z)\;,
\nonumber \\
\beta_{\mu} & \equiv & (1-\frac{4\hat{m_{l}}^{2}}{z})\;.
\end{eqnarray}

In place of $C_{7},$ one defines an effective coefficient
$C_{7}^{(0)eff}$ which is renormalization scheme independent
\cite{121}:
\begin{equation}
C_{7}^{(0)eff}(\mu _{b})=\eta ^{\frac{16}{23}}C_{7}^{(0)}(\mu _{W})+\frac{8}{%
3}(\eta ^{\frac{14}{23}}-\eta ^{\frac{16}{23}})C_{8}^{(0)}(\mu
_{W})+C_{2}^{(0)}(\mu _{W})\sum_{i=1}^{8}h_{i}\eta ^{\alpha _{i}}
\label{wilson1}
\end{equation}%
where $\eta =\frac{\alpha _{s}(\mu _{W})}{\alpha _{s}(\mu _{b})}$,
and
\begin{equation}
C_{2}^{(0)}(\mu _{W})=1,\mbox{ }C_{7}^{(0)}(\mu
_{W})=-\frac{1}{2}D^{\prime }(x_{t}),\mbox{ }C_{8}^{(0)}(\mu
_{W})=-\frac{1}{2}E^{\prime }(x_{t}); \label{wilson2}
\end{equation}%
the superscript $(0)$ stays for leading logarithm approximation,
which is not displayed in the text. Furthermore:
\begin{eqnarray}
\alpha _{1} &=&\frac{14}{23}\mbox{ \quad }\alpha _{2}=\frac{16}{23}%
\mbox{ \quad }\alpha _{3}=\frac{6}{23}\mbox{ \quad
}\alpha _{4}=-\frac{12}{23}  \nonumber \\
\alpha _{5} &=&0.4086\mbox{ \quad }\alpha _{6}=-0.4230\mbox{ \quad
}\alpha _{7}=-0.8994\mbox{ \quad }\alpha _{8}=-0.1456  \nonumber \\
h_{1} &=&2.996\mbox{ \quad }h_{2}=-1.0880\mbox{ \quad }h_{3}=-\frac{3}{7}%
\mbox{ \quad }h_{4}=-\frac{1}{14}  \nonumber \\
h_{5} &=&-0.649\mbox{ \quad }h_{6}=-0.0380\mbox{ \quad
}h_{7}=-0.0185\mbox{ \quad }h_{8}=-0.0057.  \label{wilson3}
\end{eqnarray}%
In the Naive dimensional regularization $(NDR)$ scheme  one has
\begin{equation}
C_{9}(\mu )=P_{0}^{NDR}+\frac{Y(x_{t})}{s_w^2}%
-4Z(x_{t})+P_{E}E(x_{t})  \label{wilson11}
\end{equation}%
where $P_{0}^{NDR}=2.60\pm 0.25$ \cite{121} and the last term is
numerically negligible.

$C_{10}$ is $\mu $ independent and is given by
\begin{equation}
C_{10}=-\frac{Y(x_{t})}{s_w^2}. \label{c10}
\end{equation}%
The normalization scale is fixed to $\mu =\mu _{b}\simeq 5~\rm{
GeV}$.

\null
\end{document}